\documentclass{article}
\usepackage[utf8]{inputenc}
\usepackage{graphicx}
\usepackage{subcaption}
\usepackage{graphicx}
\usepackage[
backend=biber,
style=nature,
sortlocale=de_DE,
natbib=true,
url=false, 
doi=true,
eprint=false
]{biblatex}
\title{Quantification of systematic errors in the electron density and temperature measured with Thomson scattering at \mbox{W7-X}}
\author{P.\,Nelde$^{a,b}$, G.\,Fuchert$^{a,}\footnote{Corresponding author}$, E.\,Pasch$^a$,  M.\,N.\,A.\,Beurskens$^a$,\\ S.\,A.\,Bozhenkov$^a$, K.\,J.\,Brunner$^a$, U.\,H\"ofel$^a$,\\ S.\,Kwak$^a$, J.\,Meineke$^a$, E.\,R.\,Scott$^c$, R.\,C.\,Wolf$^{a,b}$ and \mbox{W7-X}\ Team}

\date{$^a$\textit{Max-Planck-Institut f\"ur Plasmaphysik, Greifswald, Germany}\\
$^b$\textit{Technische Universit\"at Berlin, Berlin, Germany}\\
$^c$\textit{University of Wisconsin-Madison, Madison WI, U.S.A.}\\
\today}

\begin{document}

\maketitle

\begin{abstract}
The electron density and temperature profiles measured with Thomson scattering at the stellarator \mbox{Wendelstein 7-X} show features which seem to be unphysical, but so far could not be associated with any source of error considered in the data processing.
A detailed Bayesian analysis reveals that errors in the spectral calibration cannot explain the features observed in the profiles. 
Rather, it seems that small fluctuations in the laser position are sufficient to affect the profile substantially.
The impact of these fluctuations depends on the laser position itself, which, in turn, provides a method to find the optimum laser alignment in the future.
\end{abstract}

\section{Introduction}

\begin{figure}
	\centering
	\begin{subfigure}{0.49\textwidth}
		\includegraphics[width=0.95\textwidth]{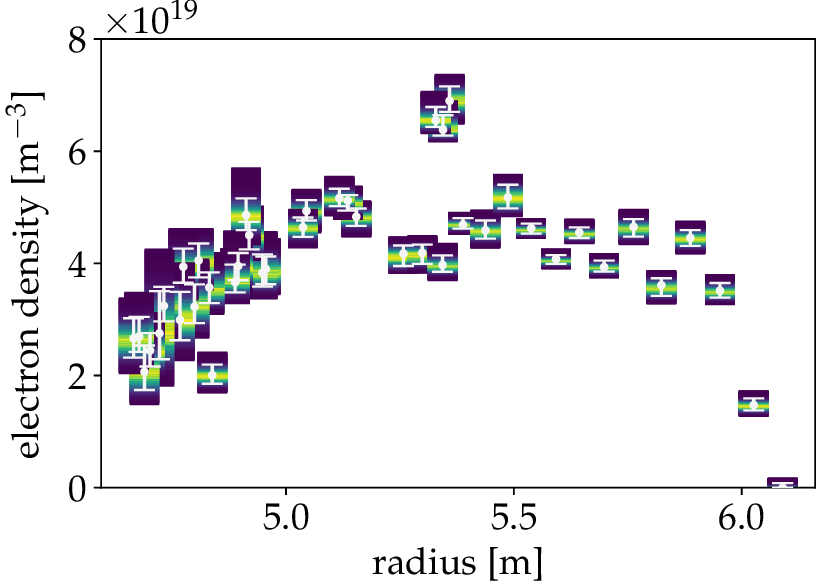}
	\end{subfigure}
	\begin{subfigure}{0.49\textwidth}
		\includegraphics[width=0.95\textwidth]{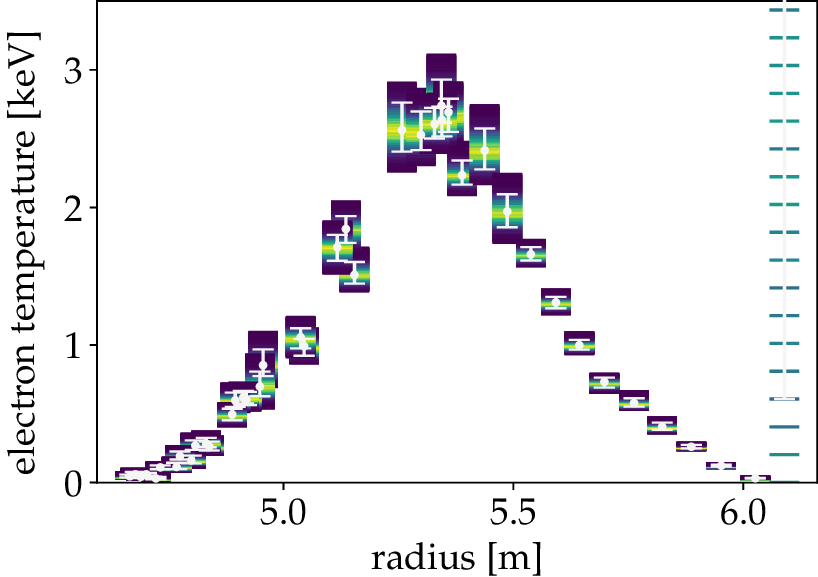}
	\end{subfigure}
	\caption[Example Thomson profile]{Typical electron density (left) and temperature (right) profiles as a function of major radius of \mbox{W7-X}. The dots represent the maximum a posteriori (the most likely) electron density/ temperature and the bars are the $95\,\%$ confidence intervals. It can be seen that if the current error estimation were to be trusted, the density profile would have a pronounced peak in the centre and an alternating behaviour at the edges.}
	\label{fig:typical_example}
	
\end{figure}

\begin{figure}
	\includegraphics[width=.95\textwidth]{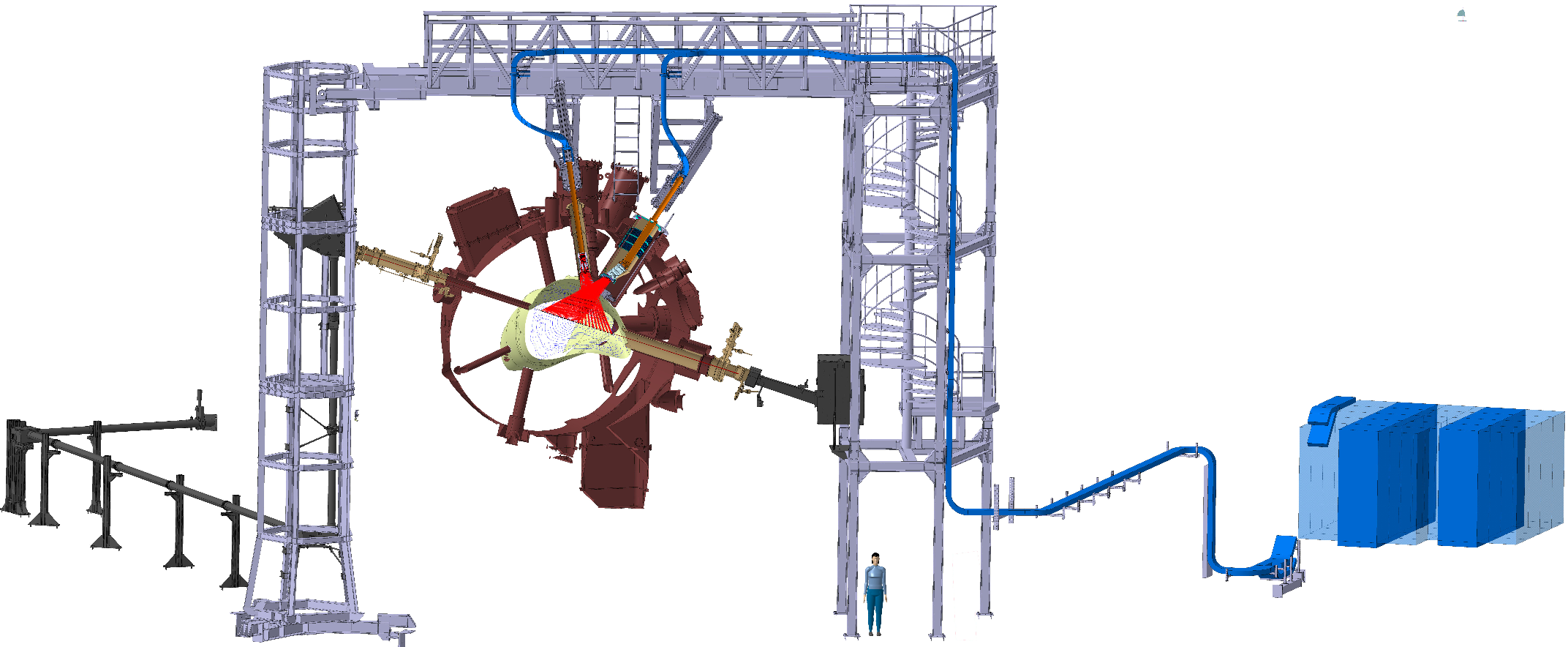}
	\caption{Schematic drawing of the Thomson scattering diagnostic at \mbox{W7-X}. The beam is guided inside the black tubing into the plasma vessel (brown). The lines of sight of the diagnostic are shown in red. The intersections of these lines with the beam path mark the measurement positions (so-called scattering volumes) inside the plasma.}
	\label{fig:setup}
\end{figure}

A reliable measurement of the electron density, $n_e$, and temperature, $T_e$, is key to gain further insight in the plasma physics of magnetic confinement devices like the stellarator \mbox{W7-X}. 
Thomson scattering is a well established method for obtaining temporally and spatially resolved $n_e$ and $T_e$ measurements \cite{pasch_2016, carlstrom_1992, narihara_2001, pasqualotto_2004, kurzan_2011}. 
The commissioning of \mbox{W7-X} is divided into separate experimental campaigns. In-between these campaigns the stellarator is being upgraded. The first experimental campaign started in 2015 and was conducted with a limiter. The second campaign started in 2017, using a test-divertor. The next campaign will add a water-cooled divertor. Each of these upgrades raises the upper limit of energy that can be coupled into the plasma. Additionally various plasma diagnostics were added and upgraded between campaigns. One diagnostic, available from the start is Thomson scattering \cite{pasch_2016, bozhenkov_2017}. However, due to the short time of operation, experience with the diagnostic is still limited.
As a consequence, not all possible error sources are fully known and could be removed or at least  quantified, yet, in order to account for them in the data evaluation. A typical set of profiles (density and temperature) at \mbox{W7-X} from the second campaign is shown in figure \ref{fig:typical_example}. 
The data is inconclusive about whether the density profile is strongly peaked or essentially flat. 
The data points in the centre appear to be elevated, but the statistical error from Bayesian analysis \cite{bozhenkov_2017} do not justify that. Hence, an additional systematic error is thought to affect the data.
Two systematic errors: spectral calibration errors and laser misalignment were suspected to be the cause of this inconsistency. With a major radius of $5.5\,\mathrm{m}$ and a plasma volume of $30\,\mathrm{m}^3$, \mbox{W7-X} is one of the largest stellarator experiments \cite{wolf_2016}. This results in the Thomson diagnostic having a beam path of about $30\,\mathrm{m}$ which is partially shown in figure \ref{fig:setup}. Along this beam path all components have to be kept stable \cite{pasch_2016}. 
For that reason, alignment was thought to be the major source of measurement errors in the experimental campaign OP1.2b (2017). The analysis presented in this paper has been performed on experimental data from that campaign. 
The paper is structured as follows: In section \ref{sec:spectral_calibration}, the influence of errors in the spectral calibration will be discussed. Following, in section \ref{sec:laser_misalignment} errors in laser alignment will be investigated. Finally, in section \ref{sec:conclusion} a conclusion and an outlook will be given, discussing strategies to mitigate such errors both at W7-X and Thomson scattering diagnostics in general.

\section{Spectral Calibration}\label{sec:spectral_calibration}

\begin{figure}
	\centering
	\begin{subfigure}[t]{0.49\textwidth}
		\includegraphics[width=0.95\textwidth]{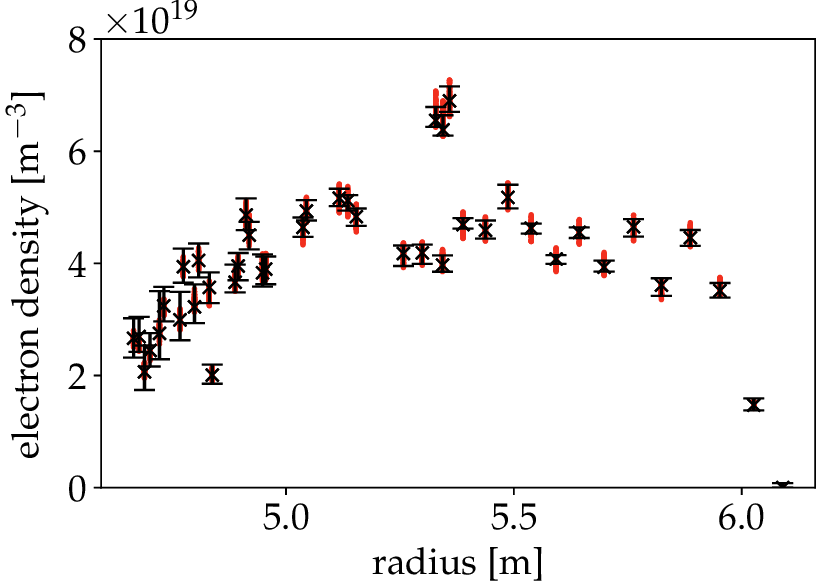}
	\end{subfigure}
	\begin{subfigure}[t]{0.49\textwidth}
		\includegraphics[width=0.95\textwidth]{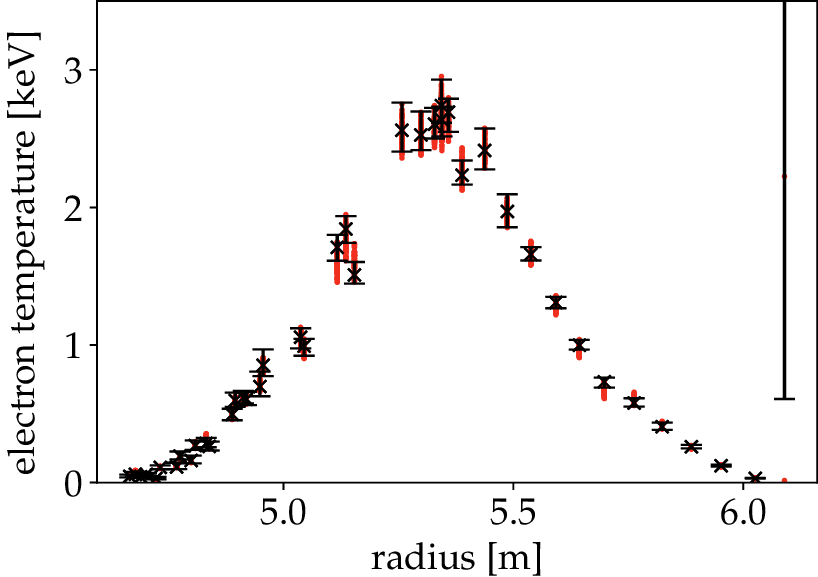}
	\end{subfigure}
	\caption[Density and temperature profiles fixed filter sets]{Shown are the electron density (left) and temperature (right) over the major radius resulting from the repeated Bayesian analysis with different sets of calibration values. The red dots mark the maximum a posteriori values of the analysis of each set. As a comparison, the black crosses indicate the experimental profiles from the standard analysis already shown in \ref{fig:typical_example}.
		It can be seen that the range of the maximum a posteriori values of the fixed filter sets are similar to the conventionally calculated confidence intervals.}
	\label{fig:hard_variation}
\end{figure}

The spectral calibration at \mbox{W7-X} is conducted with a supercontinuum pulsed light source \cite{bozhenkov_2017}.  With a monochromator a wavelength is selected and scattered diffusely with a white plate into the collection optics. The comparison of the signal behind the optics and a reference diode gives the relative calibration of the five bandpass filters used for measurement at \mbox{W7-X}. These bandpass filters are used to determine the Thomson spectrum and are henceforth called spectral channels. A $\pm 5\,\%$ deviation of the spectral calibration has been noted between several measurements in a few individual channels. In the following, the influence of these errors on the measurement of the electron density $n_e$ and temperature $T_e$ will be investigated.\\
The electron density and temperature are determined in a Bayesian way, using the Minerva framework \cite{minerva}. A typical result of the Bayesian analysis can be seen in the example figure \ref{fig:typical_example}. A strength of this kind of analysis is the possibility to calculate the whole probability distribution of the parameter in question. This distribution for $n_e$ is shown in colour-coding in the figure. The maximum a posteriori, i.e. most likely, values of $n_e$ are presented in yellow and are additionally marked by dots. The dark blue regions are regions of low probability. The bars around the dots represent the $95\,\%$ confidence intervals. \\
In this paper, the existing analysis framework was used multiple times with different sets of spectral calibrations with the individual channels varying by the $\pm5\,\%$ representing the possible error of the calibration. Each set consists of the five calibration curves, which are multiplied with $0.95$, $1.00$ or $1.05$. All possible combinations of these multipliers were tested, while keeping fixed the spectral channel absolutely calibrated by the Raman calibration \cite{bozhenkov_2017}. Mathematically speaking, this is a 4-tuple of the set $\left(0.95, 1.00, 1.05\right)$ and thus there are $3^4 = 81$ different combinations. 
\\
In figure \ref{fig:hard_variation} the $81$ maximum a posteriori values for each volume (red dots) are shown, as well as the maximum a posteriori values and error bars previously shown (black crosses) from the standard analysis (see figure \ref{fig:typical_example}).
It can be seen that both kinds of analysis are generally in good agreement. The average deviation of the mean values is around $2\,\%$, which is comparable to or even smaller than the typical statistical errors. 
Concluding, a possible error due to uncertainties in the spectral calibration is of the same order (or less) than the statistical error and could be ignored. Nevertheless, in a similar way as chosen for this analysis, it could also be included in the analysis in the future.
Additionally, the spectral calibration error can be independently checked with an alternative calibration method \cite{scott_2019}.

\section{Laser Misalignment}\label{sec:laser_misalignment}
\begin{figure}[h!]
	\centering
		\includegraphics[width=0.49\textwidth]{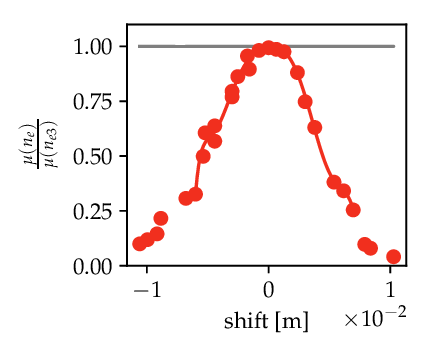}
		\includegraphics[width=0.49\textwidth]{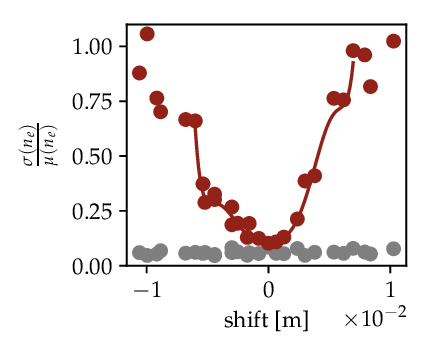}
	\caption[Density over shift one volume laser 1]{In the left the mean value and in the right the standard deviation normalised to the mean are shown as a function of the shifted beam position. A polynomial fit is applied to guide the eye.}
	\label{fig:ne_over_r_laser_1}
\end{figure}

To analyse the influence of laser misalignment, plasma experiments with nearly constant plasma parameters were conducted and two of the three lasers used for Thomson scattering were deliberately misaligned. By measuring the beam position (see appendix section \ref{sec:appendix}) and using the density measurement of the third, unchanged laser as reference, the impact on the measurement of $n_e$ can be evaluated. It is important to note that the lasers were only misaligned horizontally, which is the direction expected to have the largest impact on the measurement. In this setup, horizontal shifts are parallel to the laser axis and perpendicular to the lines of sight (i.e. out of the paper plane in figure \ref{fig:setup}).\\
To reduce the information and to gain an easy way to represent the effect of a positional shift of the laser on $n_e$, the temporal average $\mu(n_e)$ and the standard deviation $\sigma(n_e)$ were calculated for each analysed laser position. 
\\
In figure \ref{fig:ne_over_r_laser_1} (left) the mean value of $n_e$ is shown for one exemplary scattering volume located near the plasma centre. In order to account for small density variations during the experiments, $\mu(n_e)$ was normalised to the mean density of the aligned laser $\mu(n_{e3})$, which was kept steady through all the experiments (i.e. at a value of 1 both lasers agree exactly). Looking at the profile in figure \ref{fig:ne_over_r_laser_1} it can be seen that there is a significant drop in measured electron density if a laser is shifted by only a few millimetres.
\\
At the same time, due to a decreased signal-to-noise ratio the standard deviation $\sigma(n_e)$ increases, as can be seen in figure \ref{fig:ne_over_r_laser_1} (right). The behaviour is inverse to the one in the mean value. In this case, a minimum means a low scatter. 
In order to reduce the information for the following analysis, a high order polynomial fit was applied to the profiles, focusing on a good match around the extrema. 

\begin{figure}
	\centering
		\includegraphics[width=0.49\textwidth]{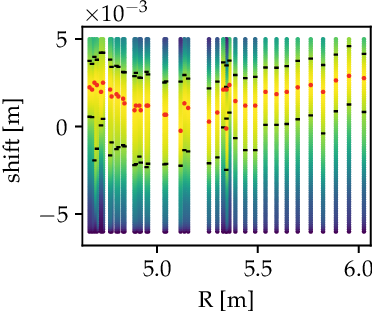}
		\includegraphics[width=0.49\textwidth]{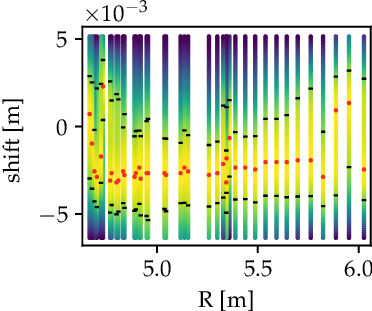}
	\caption[Density over shift all volumes]{Electron density over shift characteristics in colour scaling for all scattering volumes. Each characteristic is the result shown in figure \ref{fig:ne_over_r_laser_1} for a volume at a certain major radius $R$. A light yellow denotes a high and a dark blue a low electron density. The maxima and thus the optimal alignment positions are marked with a red dot. The black bars represent $90\,\%$ of the maximum value. In contrast to laser 1 (left), an optimal alignment position can be reached for laser 2 (right) with parallel shifts.}
	\label{fig:ne_over_r_all_volumes}
\end{figure}

The same analysis was performed for all the scattering volumes and two lasers.
In figure \ref{fig:ne_over_r_all_volumes} these polynomial fits of the mean electron densities as a function of the position shifts were converted into a colour scale (yellow being a high and blue a low value) and mapped to the respective major radius $R$ of the scattering volumes. To guide the eye the maxima are marked with red dots and $90\,\%$ of the maxima are marked with black bars. 
It can be seen that the alignment has to be treated individually for each laser. While the optimal alignment position for the second laser (right figure) can be reached for most of the scattering volumes by shifting into the negative direction, parallel shifts are not sufficient to align the first laser (left figure). As a consequence, future scans to find the optimal beam positions will include angular variations.\\

\begin{figure}[h]
	\centering
	Laser 1:
	\\
	\centering
		\includegraphics[width=0.49\textwidth]{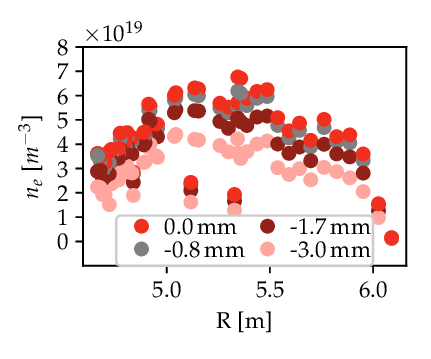}
	\hfill
		\includegraphics[width=0.49\textwidth]{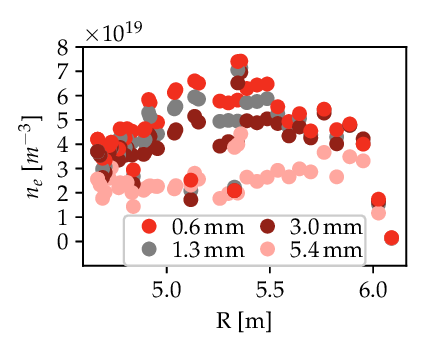}
	\\
	\centering
	Laser 2:
	\\
		\includegraphics[width=0.49\textwidth]{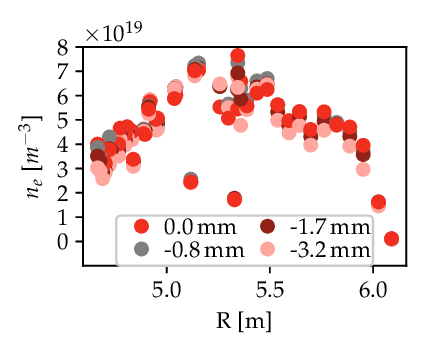}
	\hfill
		\includegraphics[width=0.49\textwidth]{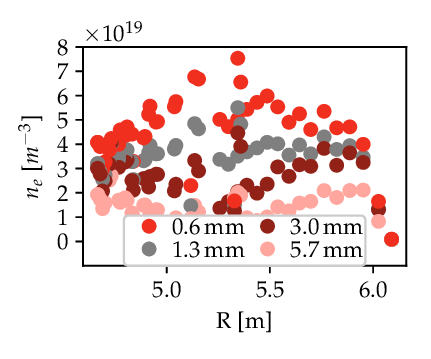}
	\caption[Density profiles with different shifts]{Electron density $n_e$ profiles as a function of the major radius $R$ for different shifts of the lasers.
		The left figures represent shifts in the negative direction and the right figures represent positive shifts. For laser 1 it can be seen that the shifts influence the profile height and shape. The same applies for positive shifts of laser 2.
		The negative shifts of laser 2 show that it is possible to align a laser such that the profiles are nearly unaffected by shifts of a few millimetres.}
	\label{fig:ne_profiles_different_shifts}
\end{figure}

To further see the impact of the induced misalignments a representative selection of averaged $n_e$ profiles are shown for different shifts of the first laser in figure \ref{fig:ne_profiles_different_shifts} (top row).
It can be seen that the shifts can change both the amplitude and the shape of the profiles, reflecting that the calibration has not been performed at the optimum position of highest intensity.
Shifts in the negative direction do not seem to change the profile shape, but the absolute measurement changes. For example a shift of $-3\,\mathrm{mm}$ reduces the measured density by one third. Roughly the same can be said about the positive shifts, with the exception of the $5.4\,\mathrm{mm}$ profile which changed significantly in shape and height. It can be concluded that with a laser shifted by $5.4\,\mathrm{mm}$ no reliable profile information can be obtained.
\\
Example profiles for the second laser are shown in figure \ref{fig:ne_profiles_different_shifts} (bottom row). Consistent with the analysis above, shifts into the negative direction seem to affect the profiles only marginally. This stable region would be the desired alignment position. 
In contrast, the positive shifts exhibit an even more pronounced change in measured density and profile shape than those of the first laser. Even a transition from a peaked profile to a hollow profile can be seen. Note that this transition is solely caused by shifting the laser and not because of changes in the plasma parameters. This is particularly problematic, because one of today's questions for \mbox{W7-X} is if the profiles are hollow, as seen in other stellarators \cite{herranz_2000} or peaked, as seen in the shown examples. If the laser alignment can also change the measured profile shape and discharges with changing plasma parameters are analysed, then answering this question becomes even more challenging. 
\\
One reason for laser 2 exhibiting such a strong dependence on shifts in the positive direction has already shown up in figure \ref{fig:ne_over_r_all_volumes}. The unshifted profile (at $\mathrm{shift}=0$) was already positioned on the positive side of the maxima. Therefore, it had to be expected that shifts in that direction would influence the profiles more strongly. This is also why the shape changes: while the outer volumes are still inside the $90\,\%$ ranges, the inner volumes are far outside these intervals. This leads to a stronger drop in the measured density in the central volumes than in the outer volumes, resulting in the observed conversion from peaked to hollow.\\
These results imply that with a similar analysis, the optimal beam alignment position can be determined before every experimental campaign to minimise the influence of the beam position on the profile shape. Such an analysis should also include angular variations, which have not been attempted in this analysis, but will added for future experimental campaigns.

\begin{figure}[h!]
	\centering
	\includegraphics[width=\textwidth]{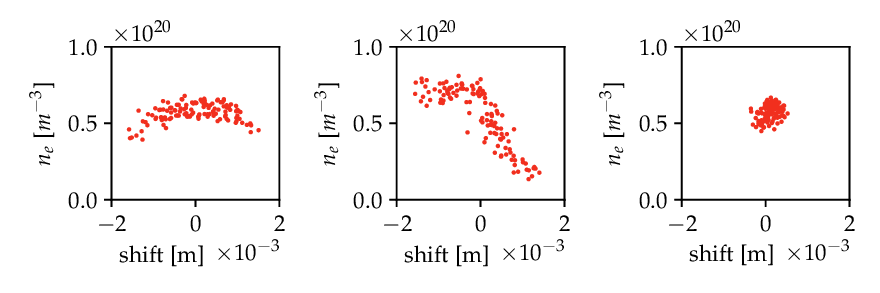}
	\caption[Density over shift non-averaged]{Non-averaged $n_e$ over shift characteristics for a volume located at major radius $R=3.692\,\mathrm{m}$. From left to right the measured data can be seen for laser 1, 2 and 3. Laser 1 measures a flat density characteristic, meaning that the same density was measured, despite positional changes between laser shots. 
		In contrast, laser 2 measures a significant drop on the positive side. 
		This shows that even if a laser is aligned for one volume, it is not necessarily aligned for all the volumes. Additionally, it can be seen that laser 3 shows less positional variation than the other 2 lasers.}
	\label{fig:single_displacement_some_volumes}
\end{figure}

Up to now, the effect of misalignment was only examined for intentional shifts and the profiles were averaged over the whole experiment programme. Because the alignment affects profile height and shape, it would be reasonable to assume that effects like beam pointing stability and mechanical vibrations of components could affect the measurements. It will be shown in the following that this is indeed the case.\\
During the analysed experiments, the measured electron density in one scattering volume should be temporally constant. Individual measurements giving evidence for that are for example the electron temperature, the diamagnetic energy (measured with diamagnetic loops) and the line integrated density (measured with an integral electron density dispersion interferometer), all of which are constant in time. In figure \ref{fig:single_displacement_some_volumes} the density for an example scattering volume is plotted over the beam position for all three lasers, with the average position being set to $0$ and deviations from that position called shifts. 
\\
At the first glance, it is apparent that the third laser is more stable, which is seen by the smaller range of shifts. It can also be seen that, the first laser has a flat density characteristic and thus is well aligned for that scattering volume. This results in a comparable jitter of $n_e$ to the one of laser 3. Laser 2, however, exhibits a significant drop of signal on the positive side, indicating an average beam position at the edge of a volume. Considering all volumes, there are cases for which laser 1 or laser 2 measure either a flat or dropping characteristic.\\ 

\begin{figure}[t]
	\centering
	\includegraphics[width=\textwidth]{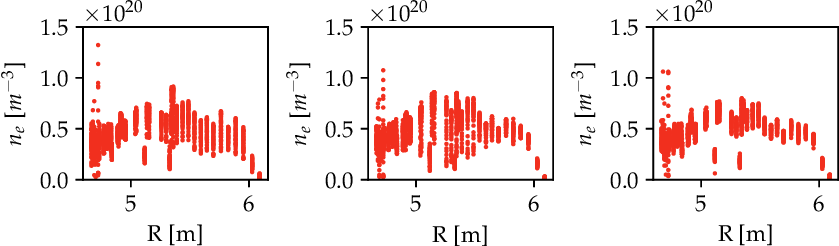}
	\caption[Density over major radius for whole experiment all lasers]{Displayed are all measured electron densities $n_e$ over the major radius $R$ for laser 1, 2 and 3 (left to right). The graphs give an impression of the error of the $n_e$ measurement.}
	\label{fig:ne_over_r_jitter}
\end{figure}

\begin{figure}[b]
	\centering
	\includegraphics[width=\textwidth]{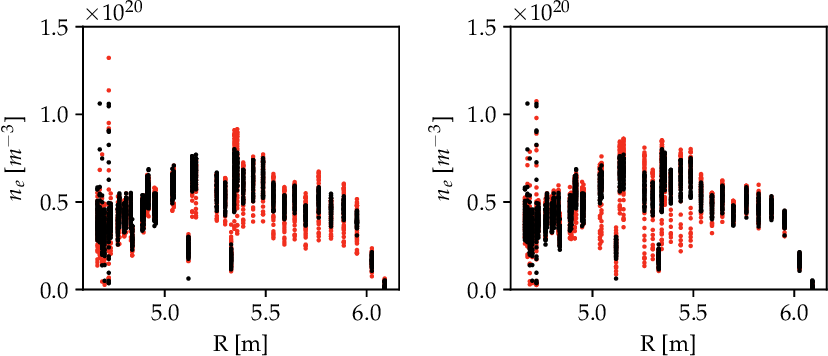}
	\caption[Density over major radius for whole experiment comparison]{Shown in red are all measured electron densities $n_e$ over the major radius $R$ for laser 1 (left) and 2 (right). For better comparison the densities measured with laser 3 are shown in black. The increase of the error caused by the stronger fluctuations of laser 1 and 2 can be seen for most volumes (major radii) by the much larger scatter of the red values. The most significant effect can be seen for the middle volumes and laser 2.}
	\label{fig:ne_over_r_jitter_compare}
\end{figure}

The characteristics with dropping edges show that the positional fluctuations introduce a significant error to the measurement. 
To give an impression of the error bars introduced by these positional fluctuations all the measured $n_e$ values are shown over the major radius $R$ in figure \ref{fig:ne_over_r_jitter}. This shows, as expected, the smallest variation in laser 3. To better compare the other two lasers with the more stable third laser, the data points of the third laser were plotted in black on top of the graphs (of figure \ref{fig:ne_over_r_jitter}) in figure \ref{fig:ne_over_r_jitter_compare}. The increase of variation can be seen by the protruding red data points. It is visible that laser 1 mostly shows more deviation for larger radii. Laser 2 shows the most deviation in the centre. There, in the margins of the possible values only because of the positional jitter either a hollow, flat or peaked $n_e$ profile seems possible.
\\
\\

The analysis revealed that the current mechanical stability of the beam path was insufficient. In the following, counter measures taken in the meantime are presented \cite{golo}. The increased positional instability of two of the lasers could be traced back to an instability in a mirror mount, which was replaced by a more robust version. Additionally, different support structures were fortified. Furthermore, in the last experimental campaign it has been observed that pressure differences between the room in which the lasers are operated and the torus hall of W7-X lead to a noticeable airflow in the beam tubes. These airflows are thought to be the main causes for mechanical vibrations along the beam path. An additional Brewster window has been installed in the beam path at the entrance to the torus hall and should reliably suppress these flows. Finally, a position resolved calibration can be performed to correct for remaining fluctuations in the beam positioning \cite{golo}.

\clearpage

\section{Conclusion and Outlook}\label{sec:conclusion}
Two systematic errors of the Thomson scattering diagnostic were investigated: the influence of errors in spectral calibration and the influence of laser misalignment. The former analysis yielded no significant influence on the derived electron density and temperature, whereas the latter analysis has shown a significant influence on the measured electron density.
\\
The presented analysis using the Bayesian Thomson model showed that the current accuracy of the spectral calibration is sufficient and a strong impact of remaining uncertainties on the derived density and temperature profiles can be excluded.
In contrast, it has been shown that the positional fluctuation of the lasers has a significant influence on the measured electron density. 
This affects each scattering volume and each laser individually and has, hence, a significant impact on both the shape and amplitude of the measured density profile.
The analysis revealed that the current mechanical stability of the beam path was insufficient and several counter measures have been implemented in the meantime. 
The next experimental campaign will show if these measures were sufficient or if they need to be taken into account during the data evaluation.

\section{Acknowledgements}
This work has been carried out within the framework of the
EUROfusion Consortium, funded by the European Union via the
Euratom Research and Training Programme (Grant Agreement No
101052200 - EUROfusion). Views and opinions expressed are
however those of the author(s) only and do not necessarily
reflect those of the European Union or the European Commission.
Neither the European Union nor the European Commission can be
held responsible for them.

\section{Appendix: Determining the beam position}
\label{sec:appendix}

\begin{figure}[h!b]
	\centering
	\includegraphics[width=\textwidth]{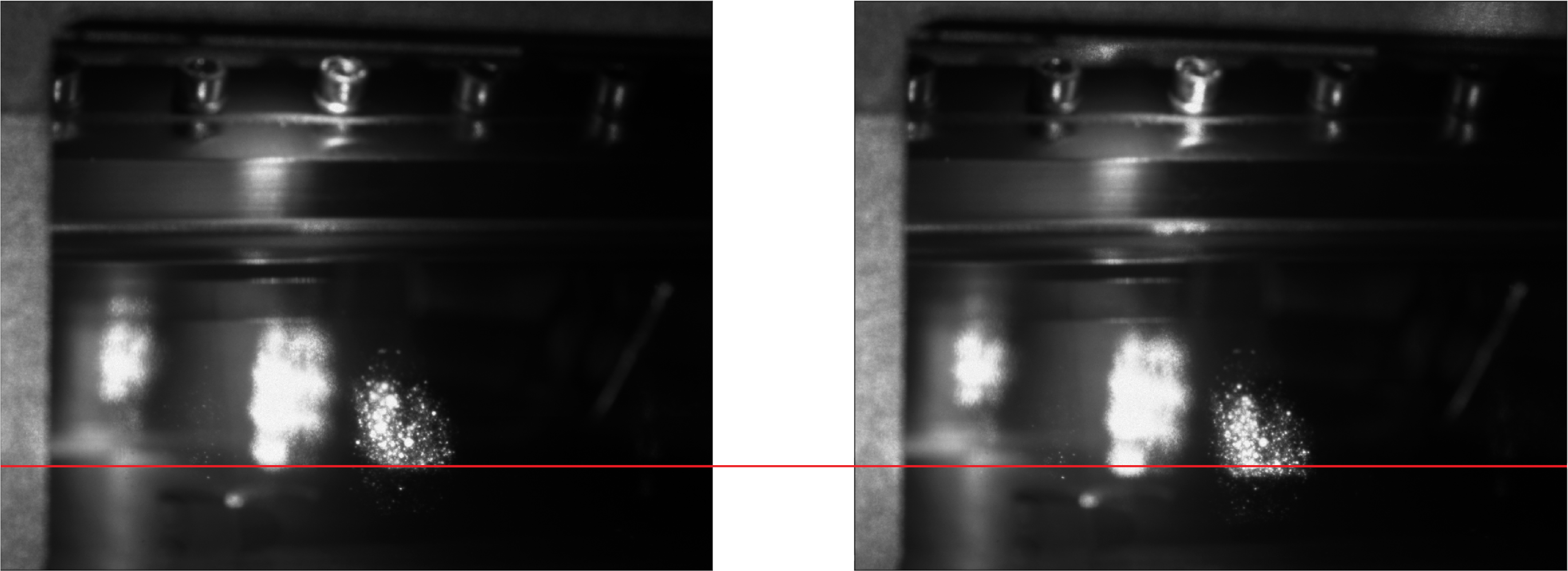}
	\caption[Laser 1 on Brewster window]{Camera images of laser 1 on the entrance Brewster window. The left image shows the laser without shift and the right image shows the laser with a $1.9\,\mathrm{mm}$ shift. The red line is a guide for the eye to see the small shift. In both images the rightmost reflection corresponds to the beam on the surface of the window being scattered by dust. The middle reflection is more consistent with the beam shape known from burning paper tests and is due to reflection on the backside of the window. This reflection is used for determining the beam position.}
	\label{fig:laser_on_brewster_ohne_Achsen}
\end{figure}

In section \ref{sec:laser_misalignment}, the beam position was used to evaluate the impact of laser misalignment on the measured electron density. In the following, it will be briefly presented how the beam position was measured.
During the experiments, the Brewster window, through which the laser passes into the plasma vessel, was monitored with a camera (located at the exit of the top left black box in figure \ref{fig:setup}). 
The beam position was determined via evaluating these cameras images. In figure \ref{fig:laser_on_brewster_ohne_Achsen} a camera image for one of the three lasers without a shift (left) and with a $1.9\,\mathrm{mm}$ shift (right) is shown. Note that a horizontal shift of the laser appears as a vertical shift in the camera view. Even with only one laser active, three light spots can be seen. The actual beam on the Brewster window corresponds to the rightmost spot and is due to scattering on dust on the windows surface. Because this speckled reflection is not well suited for determining the beam position consistently and automatically, the spot in the middle was used. This spot and the leftmost spot are resulting from reflection or multiple reflections on the backside of the Brewster window. The usage of a particular reflection should not introduce an error as long as the same reflection is used throughout the whole analysis.

\begin{figure}[h!b]
	\centering
	\includegraphics[width=\textwidth]{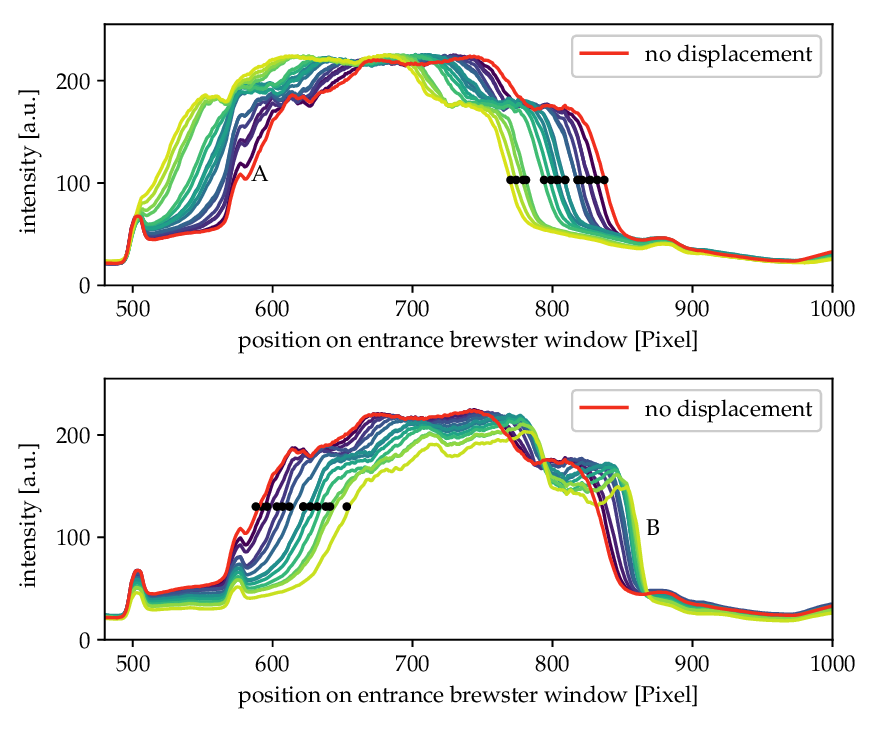}
	\caption[Laser 1 beam profiles]{Averaged beam profiles of laser 1 for each experiment sorted by shift direction. The top figure shows shifts in the left and the bottom figure shows shifts in the right direction. The red profile marks the unshifted beam. The black dots mark the points chosen to represent the beam position. At position B in the bottom image, it can be seen that the laser is being cut by an aperture in the beam path.}
	\label{fig:laser1_beam_profiles}
\end{figure}

From these camera images, the beam profiles were extracted by integrating the middle reflection along the vertical axis. The average beam position during a plasma experiment is acquired by averaging these beam profiles over the whole experiment. In figure \ref{fig:laser1_beam_profiles} the averaged beam profiles of one laser are shown for all analysed plasmas. For a better visibility, the figure was divided into negative (top) and positive (bottom) shifts of the laser. The red profile marks the laser position at the beginning of the measurement, namely the unshifted profile.

\begin{figure}[h!b]
	\centering
	\includegraphics[width=\textwidth]{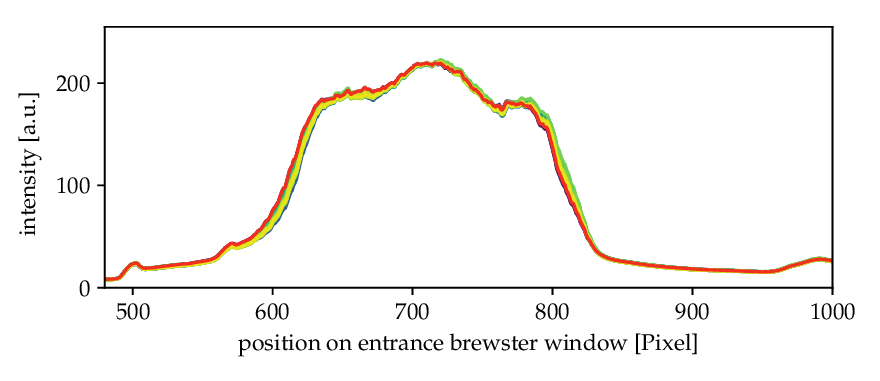}
	\caption[Laser 3 beam profiles]{Averaged beam profiles of laser 3 for each experiment. It can be seen that the average beam position could be kept stable over many experiments.}
	\label{fig:laser3_beam_profiles}
\end{figure}

Because the beam profiles do not have a pronounced maximum, a point on the profile edge was chosen to determine the beam positions. For the negative shifts (top) the right beam edge was chosen. The chosen points for each profile are marked in black. The left beam edge is not suited for this, because the shape changes with the shifts, which can be seen in the figure at the position marked \emph{A}. For positive shifts (bottom figure) the left edge was used to determine the beam positions, which are also marked with black dots. For consistency it would be better to choose the same side as before. However, this cannot be done here, because the laser is being cut off at the right edge, marked by \emph{B}. The beam position on the Brewster window can be converted into a real space scale by applying a scaling factor of $1.58\cdot10^{-4}\,\textit{m}\cdot\textit{pixel}^{-1}$.
\\
As a reference, the beam profiles of laser 3 (unshifted laser) are shown in figure \ref{fig:laser3_beam_profiles}. As in figure \ref{fig:laser1_beam_profiles} the averaged profiles are shown for every experiment. The profiles are nearly identical with the unshifted profile (red), showing that the average beam position could be kept stable over many experiments.

\clearpage

\printbibliography

\end{document}